\documentclass[twocolumn]{aastex62}

\newcommand{\lam}{$\lambda$}

\renewcommand{\ion}[2]{#1\,{\sc #2}}
\newcommand{\as}{${^\prime}{^\prime}$}
\newcommand{\ecss}{erg~cm$^{-2}$~s$^{-1}$~sr$^{-1}$}

\submitjournal{ApJ}

\shortauthors{Young et al.}

\begin{document}

\title{A Si IV/O IV electron density diagnostic for the analysis of IRIS solar spectra}

\author[0000-0001-9034-2925]{P. R. Young}
\affiliation{College of Science, George Mason University, 4400 University
  Drive, Fairfax, VA 22030, USA}
\affiliation{NASA Goddard Space Flight Center, Code 671,
  Greenbelt, MD 20771, USA}
\affiliation{Northumbria University, Newcastle Upon Tyne NE1 8ST, UK}

\author{F. P. Keenan}
\affiliation{Astrophysics Research Centre, School of Mathematics
  and Physics, Queen's University Belfast, UK}

\author{R. O. Milligan}
\affiliation{SUPA, School of Physics and Astronomy, University of Glasgow, Glasgow, G12 8QQ, UK}
\affiliation{Astrophysics Research Centre, School of Mathematics
  and Physics, Queen's University Belfast, UK}

\author{H. Peter}
\affiliation{Max-Planck Institute for Solar System Research (MPS),
  37077 G\"ottingen, Germany}

\begin{abstract}
Solar spectra of ultraviolet bursts and flare ribbons from the Interface Region Imaging Spectrograph (IRIS)  have suggested high electron densities of $>10^{12}$~cm$^{-3}$ at transition region temperatures of 0.1~MK, based on large intensity ratios of \ion{Si}{iv} \lam1402.77 to \ion{O}{iv} \lam1401.16. In this work a rare observation of the weak \ion{O}{iv} \lam1343.51 line is reported from an X-class flare that peaked at 21:41~UT  on 2014 October 24. This line is used to develop a theoretical prediction of the \ion{Si}{iv} \lam1402.77 to \ion{O}{iv} \lam1401.16 ratio as a function of density that is recommended to be used in the high density regime. The method makes use of new pressure-dependent ionization fractions that take account of the suppression of dielectronic recombination at high densities. It is applied to two sequences of flare kernel observations from the October 24 flare. The first shows densities that vary between $3\times 10^{12}$ to $3 \times 10^{13}$~cm$^{-3}$ over a seven minute period, while the second location shows stable density values of around $2\times 10^{12}$~cm$^{-3}$ over a three minute period.
\end{abstract}


\section{Introduction}

The Interface Region Imaging Spectrograph \citep[IRIS;][]{iris} observes four of the five \ion{O}{iv} intercombination lines between 1399 and 1405~\AA, and the two resonance lines of \ion{Si}{iv} at 1393.76 and 1402.77~\AA, which together diagnose the solar transition region around 0.1~MK. The \ion{O}{iv} lines are well known to form density diagnostics that are valuable for solar and stellar spectroscopists \citep[e.g.,][]{keenan02}. However, at high electron number densities of 
$N_{\rm e}> 10^{12}$~cm$^{-3}$,  the \ion{O}{iv} diagnostics lose
sensitivity and so are no longer useful. It was noted by \citet{feldman77} that the \ion{O}{iv} lines become very weak relative to the \ion{Si}{iv} lines in flare spectra, which was interpreted as collisional de-excitation becoming the
dominant de-population method for the \ion{O}{iv} lines' upper levels at high densities, thus
reducing their photon yields in comparison to the \ion{Si}{iv} resonance lines. The ratio of a \ion{Si}{iv} line to a \ion{O}{iv} line can thus be used as a density diagnostic in the regime $N_{\rm e}> 10^{12}$~cm$^{-3}$  that is not otherwise accessible by IRIS. \citet{feldman77} constructed a method for interpreting the ratio that yielded a value of
$>10^{13}$~cm$^{-3}$ for one \emph{Skylab} solar flare
observation, and similar results were presented by later authors---see \citet{doschek16} for a review of results from the \emph{Skylab} and Solar Maximum Mission experiments.

After the launch of IRIS in 2013 there has been renewed interest in the use of \ion{Si}{iv} and \ion{O}{iv} to derive densities,
particularly with regard to intense transition region bursts that
occur within active regions. These ultraviolet (UV) bursts demonstrate very strong
\ion{Si}{iv} emission and also weak or non-existent \ion{O}{iv}
emission. 
\citet{peter14} presented four examples that they named ``bombs", and which exhibited lower limits of 50--350 for the \ion{Si}{iv} \lam1402.77 to \ion{O}{iv} \lam1401.16 ratio, which compare with a value of around 4 in quiet Sun conditions \citep{doschek01}. \citet{2016ApJ...824...96T} studied ten UV bursts and found ratios ranging from 32 to 498 for eight of the events, but also two which had low ratios of 7. They inferred that the ratios (and thus densities) are sensitive to the heights of formation of the bursts.
To convert measured ratios to densities, \citet{peter14} created a constant pressure model for the line emissivities, and this yielded densities of $\gtrapprox 10^{13}$~cm$^{-3}$ for the four UV bursts studied in this work. This method was also applied by \citet{2015ApJ...810...38K},
\citet{2015ApJ...811...48Y} and \citet{2015ApJ...809...82G} to other UV bursts, and similar high densities were found.
The high
densities 
found for the UV bursts were
cited by \citet{peter14} as evidence for their formation in the temperature minimum
region (500~km above the solar surface) or lower.
However, \citet{2015ApJ...808..116J} questioned the \citet{peter14} result,
citing  uncertainties in element abundances, temperature
structure and atomic data, and argued that the UV bursts are likely formed in
a lower density regime that is higher in the atmosphere. Despite these
concerns, a single density--ratio relationship for \ion{Si}{iv}
and \ion{O}{iv} is very appealing due to the paucity of other IRIS 
diagnostics in the high density regime, as demonstrated by the use of
the method in \citet{2015ApJ...810...38K},
\citet{2015ApJ...811...48Y}, \citet{2015ApJ...809...82G} and
\citet{2016A&A...594A..64P}. For this reason, we re-visit the
\ion{Si}{iv}-to-\ion{O}{iv} 
ratio method here.

\citet{2015ApJ...808..116J} recommended that the \ion{O}{iv} allowed multiplet
around 1340~\AA\ (two lines at 1338.61 and 1343.51~\AA) be included in
the IRIS analysis, but these wavelengths are often not downloaded due to IRIS telemetry restrictions. Even when the wavelengths are downloaded, the lines are usually too weak to observe. For the present work we utilize a flare data-set where \lam1343.51 is observed and  automatic exposure control reduces the exposure time by a factor of more than 30, thus enabling reliable intensity estimates for the \ion{O}{iv} and \ion{Si}{iv} lines. This allows the \ion{Si}{iv}--\ion{O}{iv} density diagnostic method to be investigated, and also for the method to be benchmarked against the usual intercombination line density diagnostic.

Section~\ref{sect.atom} describes the atomic data and emission line
modeling technique used in the present work.  A method for creating a \ion{Si}{iv} to \ion{O}{iv} density diagnostic based on the quiet Sun differential emission measure (QS-DEM) is presented in Section~\ref{sect.qsdem}. The observational data-set is presented in Section~\ref{sect.obs}, and densities derived using the QS-DEM method.
A modified method---referred to as the ``log--linear DEM" method---that makes use of the \lam1343.51 line is then developed that we recommend to use in the high density regime ($>10^{12}$~cm$^{-3}$). This method is then applied to derive densities for a series of flare kernel measurements in Sect.~\ref{sect.kernels}, and a final summary is given in Section~\ref{sect.summary}.

\section{Atomic data and line modeling}\label{sect.atom}

The emission lines considered here are listed in
Table~\ref{tbl.lines}. Those of \ion{Si}{iv}  are strong resonance
transitions for which the intensity  increases as $N_{\rm e}^2$, while the
\ion{O}{iv} and \ion{S}{iv} lines are 
intercombination transitions  emitted from metastable
levels. At low densities ($\log\,N_{\rm e} \lesssim 10$) the
\ion{O}{iv} and \ion{S}{iv} lines
behave as allowed transitions. However, eventually the density
becomes high enough that electron collisional de-excitation dominates
radiative decay in de-populating the levels. The intensities of the  lines then increase only as $N_{\rm e}$.
This explains why the intercombination
lines would be expected to become much weaker than the \ion{Si}{iv}
allowed transitions at high density, the basis for the \ion{Si}{iv}--\ion{O}{iv}
density diagnostic.

\begin{deluxetable*}{ccccc}
\tablecaption{Emission lines studied in the present work.\label{tbl.lines}}
\tablehead{
  \colhead{Ion} &
  \colhead{$\log\,(T_{\rm max}/{\rm K})$} &
  \colhead{Wavelength/\AA} &
  \colhead{Transition} &
  \colhead{$\log\,(T_{\rm mem}/{\rm K})$}
}
\startdata
\ion{O}{iv} & 5.17&1343.51 & $2s2p^2$ $^2P_{3/2}$ -- $2p^3$ $^2D_{5/2}$  &5.15 \\
            && 1399.78 & $2s^22p$ $^2P_{1/2}$ -- $2s2p^2$ $^4P_{1/2}$ &5.09\\
            && 1401.16 & $2s^22p$ $^2P_{3/2}$ -- $2s2p^2$ $^4P_{5/2}$ &5.09\\
                        && 1404.78 & $2s^22p$ $^2P_{3/2}$ -- $2s2p^2$ $^4P_{3/2}$ &5.09\\
\noalign{\smallskip}
\ion{Si}{iv} &4.88 & 1393.76 & $3s$ $^2S_{1/2}$ -- $3p$ $^2P_{3/2}$ & 4.85 \\
             && 1402.77 & $3s$ $^2S_{1/2}$ -- $3p$ $^2P_{3/2}$ & 4.85 \\
\noalign{\smallskip}
\ion{S}{iv} & 4.98 & 1404.83 & $3s^23p$ $^2P_{1/2}$ -- $3s3p^2$ $^4P_{1/2}$&4.89\\
\enddata
\end{deluxetable*}

The atomic models for  \ion{O}{iv}, \ion{Si}{iv} and \ion{S}{iv}
are taken from 
version
8.0 of the CHIANTI database
\citep{chianti8,2016JPhB...49g4009Y}. For \ion{Si}{iv} the observed energy
levels are from the 
NIST database\footnote{\url{https://www.nist.gov/pml/atomic-spectra-database}.}, with radiative decay rates and electron collision
strengths from
\citet{2009JPhB...42v5002L} and \citet{2009A&A...500.1263L}. Energy
levels for \ion{O}{iv} are from \citet{feucht97} and the NIST
database; for the three $2s2p$ $^4P_J$ levels that give rise to lines
in the IRIS wavebands the energies are derived from the wavelengths of
\citet{sandlin86}---see \citet{2016A&A...594A..64P}. Radiative decay rates and electron collision
strengths are from \citet{liang-b}. Energy levels for \ion{S}{iv} 
are from the NIST database, and radiative decay rates are from
\citet{hibbert02}, \citet{tayal99}, \citet{johnson86} and an
unpublished calculation of P.R.~Young. Electron collision strengths
are from \citet{tayal00}. 

Equilibrium ionization fractions as a function of temperature are
distributed with CHIANTI, and these are computed in the so-called
``zero-density'' approximation, assuming that ionization and
recombination occur only from the ions' ground states, and that the
rates are independent of density. However, it is known that
dielectronic recombination rates become suppressed at high density, and
so in the present work we modify the rates used in CHIANTI with the
\citet{nikolic13} suppression factors. These lead to  density-dependent ion fractions, although for our work we derive ion fraction
tables as a function of pressure. Further details on the implementation are given in \citet{2018arXiv180105886Y}.

In Table~\ref{tbl.lines}   the temperature of maximum
ionization, $T_{\rm max}$, of an ion and the temperature of maximum
emission, $T_{\rm mem}$, of an emission line are given. We define the former to be the
temperature at which the zero-density ionization fraction curve of the ion peaks. The latter is the temperature at which the contribution function for a
specific emission line peaks. We define $T_{\rm mem}$ here by assuming a density of $10^{12}$~cm$^{-3}$ and using the new ionization fraction curves computed at this density. The consequence of DR suppression is to push the ions to lower temperatures, and hence the values of $T_{\rm mem}$ are lower than those of  $T_{\rm max}$. We also note that 
the $T_{\rm mem}$ value
for \lam1343.51 is significantly higher than for the other \ion{O}{iv} lines, which is a 
consequence of it being a high excitation line.

Flare kernels usually show sudden large intensity increases and thus ionization equilibrium may be a questionable assumption. However, we note 
that previous authors have demonstrated that non-equilibrium effects are only important at relatively low densities and over short timescales. 
For example, \citet{noci89} considered 
non-equilibrium ionization for carbon ions in a coronal loop model that
featured a steady siphon flow, and at loop densities
$> 4\times 10^9$~cm$^{-3}$ (significantly lower than the
$>10^{11}$~cm$^{-3}$ 
values considered here) the effects were small. \citet{2013A&A...557L...9D}
modeled the \ion{O}{iv} and \ion{Si}{iv} emission lines with an atomic
model that fully incorporated density effects in the ion balance, and
found that \ion{O}{iv} to \ion{Si}{iv} line ratios in a
transiently-heated plasma reach their ionization equilibrium values
within 10~seconds for a density of $10^{10}$~cm$^{-3}$. For higher
densities, this time would be reduced further. 
\citet{2013ApJ...767...43O} applied a sophisticated 3D radiative magnetohydrodynamic model to the study of the \ion{O}{iv} \lam1401.16/1404.78 density diagnostic and found that non-equilibrium ionization results in \ion{O}{iv} being formed over a wider range of temperatures than expected from the equilibrium ionization fractions. The low temperature contributions correspond to low heights in the atmosphere and thus higher densities. Therefore the simulated ratios tend to correspond to higher densities than for the equilibrium case, and they do not represent the density at the $T_{\rm max}$ of the ion. 
The range of densities over which \ion{O}{iv} is sensitive to in this model is $\log\,N_{\rm e}=8.2$--10.5, and we emphasize again that this is much lower than the densities considered in the present work. If a high density ``knot" of plasma were present in this model, we would expect it to be in ionization equilibrium, and equilibrium diagnostics would apply. 
To summarize this discussion, we are 
justified in assuming ionization equilibrium for the high density features studied in the present work.

We also assume that the measured line intensities come from a single structure when interpreting the line ratios. The effect of multiple structures with different densities (or pressures) on the interpretation of density diagnostics was considered by \citet{1984ApJ...279..446D} and \citet{1997ApJ...475..275J}. In particular, the derived density of a high density structure can be lower than the actual density if there is a significant amount of low density plasma in the instrument's spatial resolving element. However, we consider this a relatively small effect for IRIS observations of flare kernels and bursts due to the high spatial resolution of IRIS and the fact that the densities of these features are two to three orders of magnitude higher than typical background active region plasma.

A further approximation used here is that particle distributions are
described by Maxwellians. The effects of non-Maxwellians---in
particular, $\kappa$-distributions---on the  ratios of \ion{O}{iv}
to \ion{Si}{iv} lines  were studied by \citet{dudik14}, who
found that the \ion{O}{iv} lines can be  suppressed relative to
\ion{Si}{iv} by more than an order of magnitude as the $\kappa$ index
decreases from 10 to 2 (i.e., becoming less Maxwellian).
However, for the high
densities considered in the present work, we do not expect
non-Maxwellian distributions to be maintained for periods long enough
to affect the line intensities to any significant extent.

Adopting the above  approximations, CHIANTI is used to model the intensities
of the emission lines, and the method follows that of the companion
paper \citet{2018arXiv180105886Y} in writing a line intensity, $I$, as a
sum of isothermal components. We assume a given structure has constant pressure, $P=N_{\rm e}T$, which is motivated by the
consideration that the plasma will be trapped along field lines, and
that the thickness of the transition region is significantly smaller
than the pressure scale height at transition region temperatures. Hence  $I$ is written as a function of pressure:
\begin{equation}\label{eq.int}
I(P) = \epsilon({\rm X})\sum_k  {G(T_k,P) P^2 H(T_k) h_k \over T_k^2} 
\end{equation}
where $\epsilon({\rm X})$ is the abundance of the emitting element X, $G(T,P)$ is the contribution function,  $H(T)$ is the ratio of hydrogen to free electrons ($N_{\rm H}/N_{\rm e}$), and $h$ is the emitting column depth of the
plasma. We calculate $G$ using  the CHIANTI software routine \verb|gofnt.pro|, and $H$ with the routine \verb|proton_dens.pro|,
which takes into account the element
abundance and ionization fraction files. 
The sum is performed over a grid of temperatures, $T_k$, that have a spacing of 0.05~dex  in $\log\,T$ space. 

The differential emission measure (DEM), $\phi(T)$, is related to $h$ by the expression
\begin{equation}\label{eq.phi}
\phi = { P^2H(T) h  \over 0.115 T^3}.
\end{equation}
Note that the numerical factor in the denominator comes from $\delta T=T\ln\,10\,\delta(\log\,T)=0.115T$, where $\delta T$ is the size of the temperature bin.

The assumption of constant pressure for the emission line modeling has the potential to cause confusion when referring to density diagnostics formed from two ions with different formation temperatures. For example, if the pressure is $\log\,(P/{\rm K~cm}^{-3})=15$ then the electron density at the $T_{\rm max}$ value of \ion{Si}{iv} is $\log\,(N_{\rm e}/{\rm cm}^{-3})=10.12$, but for \ion{O}{iv} the density is $\log\,(N_{\rm e}/{\rm cm}^{-3})=9.83$. In the following text we therefore refer to line ratio pressure diagnostics rather than density diagnostics. Where we refer to density, we will give the quantity $N_{\rm e}^{\rm Si\,IV}$, which is the electron density at a temperature of $\log\,(T/{\rm K})=4.88$, the $T_{\rm max}$ of \ion{Si}{iv}.

\section{Emission line modeling}\label{sect.qsdem}

In this section we derive pressure--ratio curves for various
combinations of \ion{Si}{iv} and \ion{O}{iv} lines using a 
method that we refer to as the ``QS-DEM method'', as it is based on the
quiet Sun DEM. Each curve allows the user to convert a single  observed line ratio directly to a pressure, and it is useful for 
the regime in which the \ion{O}{iv} lines are no longer density sensitive ($N_{\rm e}\ge
10^{12}$~cm$^{-3}$), and when \ion{O}{iv} \lam1343.5 is
unavailable. 

Due to the separation in temperature of \ion{Si}{iv} and \ion{O}{iv} it is necessary to make an assumption about the plasma temperature structure. Here,  the DEM in the vicinity  of the \ion{Si}{iv} and \ion{O}{iv} lines is assumed to take a  universal shape that can be applied to any feature. This assumption fixes the relative values of $h_k$ in Eq.~\ref{eq.int}, and means that the ratio of any two emission lines depends solely on pressure.

Standard quiet Sun (QS) DEMs are available in the literature with perhaps the
most well-known being the one distributed with the CHIANTI database. However, for our
work we wish to systematically apply the pressure sensitive ion balance
described in the previous section, which  means generating  the DEM curve with these results. For this purpose we make use of results  from the
companion paper \citep{2018arXiv180105886Y}  that derived abundance ratios of magnesium
to neon (Mg/Ne) and neon to oxygen (Ne/O) in the average quiet
Sun. The latter is the most relevant, as the temperature range
of the ions overlaps with the \ion{Si}{iv} and \ion{O}{iv} ions used
here.

\citet{2018arXiv180105886Y}  computed intensities using Eq.~\ref{eq.int} by assuming that $h(T)$ takes a bilinear form such that there are three node points at $\log\,T=4.5$, 5.2 and 5.8, with values $h=p_1$, $p_2$ and $p_3$. The other values of $h$ on the temperature grid are determined by linear interpolation in $\log\,T$--$\log\,h$ space between $p_1$ and $p_2$, and $p_2$ and $p_3$. Thus the function $h(T)$ is defined entirely by $p_{1,2,3}$. A $\chi^2$ minimization was then performed to reproduce the intensities of the six observed oxygen and neon line intensities by varying $p_{1,2,3}$ and the Ne/O relative abundance. A standard quiet Sun pressure of $10^{14.5}$~K~cm$^{-3}$ was used.

\begin{figure}[t]
\plotone{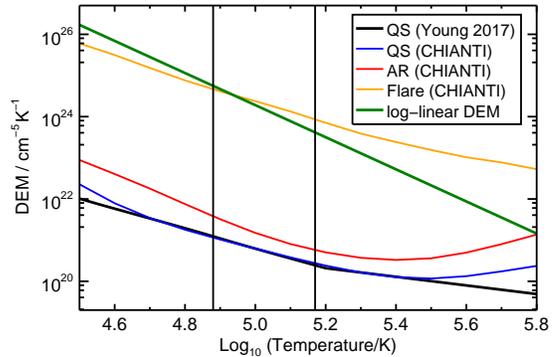}
\caption{A comparison of the quiet Sun DEM used in the present work (black line) with the quiet Sun, active region and flare (blue, red and orange, respectively) DEMs available in CHIANTI, and the log--linear DEM (green) derived in Section~\ref{sect.linlog}. The vertical black lines show the $T_{\rm max}$ values of \ion{Si}{iv} and \ion{O}{iv}.}
\label{fig.dem}
\end{figure}

The above procedure was applied to 24 quiet Sun CDS datasets obtained over the period 1996--1998. For the present work, we averaged the 24 sets of derived $p_{1,2,3}$ values, giving $p_1=3.93$, $p_2=11.2$ and $p_3=166$~km. These values essentially define the quiet Sun DEM used in this section (noting the relation between $\phi$ and $h$ from Eq.~\ref{eq.phi}), and the DEM values are plotted in Figure~\ref{fig.dem}.

For the element abundances in Eq.~\ref{eq.int} we adopt  the  photospheric oxygen and silicon  abundances of \citet{caffau11} and \citet{lodders09}, respectively, but multiply the latter by a factor 1.6, reflecting the enhanced Mg/Ne abundance ratio found in the quiet Sun transition region by \citet{2018arXiv180105886Y}. Silicon has a low first ionization potential (FIP), like magnesium, and so would be expected to be similarly enhanced compared to the high FIP element oxygen.

With the abundances and the column depths defined, we use a quiet Sun pressure of $\log\,(P/{\rm K~cm}^{-3})=14.5$ to yield intensities for \ion{Si}{iv} \lam1402.77 and \ion{O}{iv} \lam1401.16 of 22.0 and 18.4~\ecss, respectively. \citet{doschek01} measured a \lam1401.16/\lam1402.77 ratio of $0.267\pm 0.050$ from quiet Sun spectra obtained by SUMER, which is a factor of 3.13 larger than the ratio predicted from the DEM. 
We consider this to be an \emph{empirical correction factor} that is necessary to apply to the \ion{Si}{iv} lines in order to address a well-known problem first highlighted by \citet{1972ApJ...178..527D}.
Specifically, the observed intensities of lines from the
lithium- and sodium-like isoelectronic sequences are usually 
stronger than one would expect based on the emission measures from other sequences formed at the same temperature. 
This is why we  apply the correction factor to
the silicon lines and not those of oxygen.

Normalizing the ratio in this way was first undertaken by \citet{feldman77}
for the analysis of \emph{Skylab} spectra. It was also employed by
\citet{doschek16} in a recent analysis of IRIS spectra, but not
for other IRIS work 
\citep{peter14,2015ApJ...810...38K,2015ApJ...811...48Y,
  2015ApJ...809...82G,2016A&A...594A..64P}. 

\begin{figure}[t]
\epsscale{1.0}
\plotone{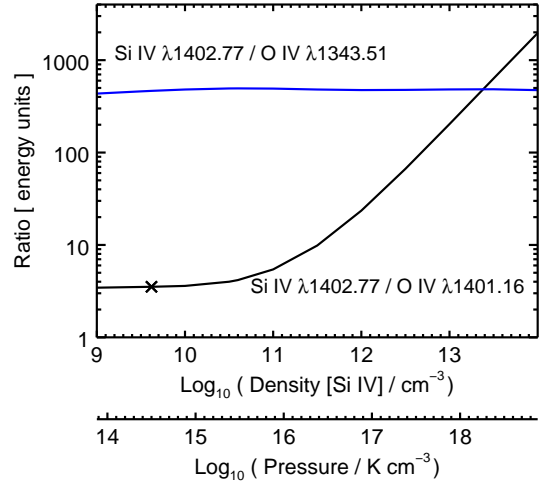}
\caption{\ion{Si}{iv}/\ion{O}{iv} ratio--density curves predicted from
the QS-DEM method. These are plotted as a function of the density at $\log\,T=4.88$.}
\label{fig.ratios}
\end{figure}

With the empirical correction factor defined, the intensities of the
\ion{Si}{iv}, \ion{O}{iv} and \ion{S}{iv} lines are computed with Eq.~\ref{eq.int} for a range of pressures using the $h_k$ values from the quiet Sun analysis. We then take ratios of a selection of lines and present the results in Table~\ref{tbl.data}, with two of the ratios displayed graphically  in
Figure~\ref{fig.ratios}. The \lam1393.76/\lam1401.16 ratio can be directly compared to that given by \citet{peter14}. Ratio values of 10 and 700 correspond to densities $\log\,N_{\rm e}^{\rm Si\,IV}=11.7$ and 13.3 in their work, compared to 10.8 and 13.2 here.
The \lam1402.77/1343.51 ratio is insensitive to pressure, with an average value of 514 over a range of five orders of magnitude in pressure. This large ratio shows why the \lam1343.51 is difficult to measure in IRIS observations: the noise level of the spectra is around 10~DN (data numbers), and the maximum signal is about 16\,000~DN before saturating. Thus  \lam1343.51 only becomes measurable when \lam1402.77 is close to saturation.

Finally in this section we return to two of the assumptions that were made for the modeling described above; specifically the element abundances and shape of the DEM curve. 
Our method assumes that the events have a Si/O abundance ratio that is consistent with the small FIP enhancement factor found  by 
\citet{2018arXiv180105886Y}. The two IRIS features that we know have high densities are flare kernels (an example is studied here) and UV bursts, which are compact, intense brightenings seen in active regions but not related to flares \citep[the bomb events studied by][are examples]{peter14}. Both of these features are highly impulsive, and we note that
\citet{2016ApJ...824...56W} recently found that impulsive active
region heating events show close-to photospheric abundances.

The suitability of using a quiet Sun DEM for UV bursts was discussed in
detail by \citet{doschek16} with reference to \emph{Skylab} spectra
from which a much wider range of emission lines was available. They
concluded that the shape of the QS DEM in the vicinity of the \ion{Si}{iv}
and \ion{O}{iv} ions is approximately the same in active regions in
an average sense, and 
that the variations in the \ion{Si}{iv}/\ion{O}{iv} ratio
across an active region are driven largely by density variations
rather than those in temperature or DEM. This is  borne out by a comparison of the \citet{2018arXiv180105886Y}  quiet Sun DEM with the quiet Sun, active region and flare DEMs available in the CHIANTI database (Figure~\ref{fig.dem}). Without additional
ions in the IRIS datasets to derive DEMs for UV bursts and flare kernels, we feel justified in using the quiet Sun DEM here.

\begin{deluxetable*}{cccccccccccc}
\tablecaption{Theoretical ratios in energy units from the QS-DEM method. \label{tbl.data}}
\tablehead{
   & \multicolumn{11}{c}{$\log\,(N_{\rm e}^{\rm Si\,IV}/{\rm
      cm}^{-3})$} \\
\cline{2-12}
  Ratio &
  9.0 & 9.5 & 10.0 & 10.5 & 11.0 & 11.5 &
  12.0 & 12.5 & 13.0 & 13.5 & 14.0 
}
\startdata
\sidehead{\ion{Si}{iv}/\ion{O}{iv} ratios}
\lam1393.76/\lam1401.16 &
    7.45&    7.59&    7.79&    8.64&   11.74&   21.29&   50.72&  143.6&  437.8& 1362.6& 4221.3\\
\lam1402.77/\lam1401.16 &
    3.73&    3.80&    3.90&    4.33&    5.88&   10.67&   25.42&   71.99&  219.5&  682.9& 2115.1\\
\lam1402.77/\lam1343.51 &
  469.4&  496.3&  520.1&  534.1&  532.3&  520.1&  512.9&  514.9&  520.4&  522.3&  510.8\\
\sidehead{\ion{O}{iv}/\ion{O}{iv} ratios}
\lam1401.16/\lam1343.51 &
 125.8& 130.5& 133.3& 123.3&  90.45&  48.75&  20.17&   7.15&   2.37&   0.765&   0.241\\
\lam1399.78/\lam1401.16 &
   0.171&   0.177&   0.193&   0.228&   0.288&   0.354&   0.396&   0.415&   0.422&   0.424&   0.425\\
\lam1404.78/\lam1401.16 &
   0.560&   0.505&   0.403&   0.288&   0.214&   0.182&   0.171&   0.168&   0.167&   0.166&   0.166\\
\sidehead{\ion{S}{iv}/\ion{O}{iv} ratios}
\lam1404.83/\lam1401.16 &
   0.025&   0.025&   0.026&   0.029&   0.041&   0.075&   0.148&   0.241&   0.308&   0.339&   0.351\\
\enddata
\end{deluxetable*}

\section{Observations and measurements}\label{sect.obs}

IRIS is described in detail by \citet{iris}, and hence here we only briefly summarize the important features relevant to the present work.  A single telescope feeds a spectrograph and slitjaw imager (SJI), allowing simultaneous imaging and slit spectroscopy. Spectra are obtained in three wavelength bands, and the lines studied in the present work are from the far ultraviolet wavebands 1331.7--1358.4 and 1389.0--1407.0~\AA\ (FUV1 and FUV2, respectively). The spectral resolution is about 53\,000 and the spatial resolution is 0.33\as.

\begin{figure*}[t]
\epsscale{1.0}
\plotone{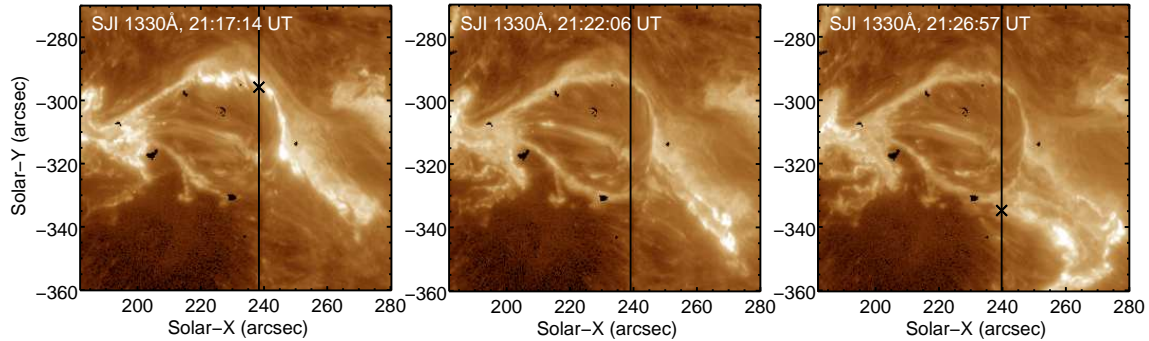}
\caption{Images of the flare ribbon  obtained with the IRIS SJI 1330~\AA\ filter. The location of the IRIS slit is marked with a vertical black line, and the two spatial locations studied are marked with crosses in the left and right panels.}
\label{fig.sji}
\end{figure*}

The observations are from active region AR 12192, one of the best known of Solar Cycle 24 due to its very large sunspot, and the high number of confined flares it produced during its transit across the solar disk during 2014 October 17--30. The IRIS observation that began at 20:52~UT on October 24 is used  and it consisted of a 6.75~hour sit-and-stare dataset that captured a long-duration X3.1 flare, which peaked at 21:41~UT. Our interest here lies with the rise phase of the flare when the IRIS slit lay across one of the developing ribbons, and three SJI images of the ribbon are shown in Figure~\ref{fig.sji}. This ribbon had a complex shape and time evolution, and it produced intense flare kernels in \ion{Si}{iv} at two locations, namely $Y=-295$\as\ during 21:11 to 21:18~UT, and $Y=-330$\as\ during 21:25 to 21:28~UT. The observational sequence began with 15~second exposure times, but at 21:11:18~UT exposure control was triggered, resulting in sub-second exposures for the period up to 21:36:41~UT. These enabled unsaturated profiles of the strong \ion{Si}{iv} \lam1402.77 line to be obtained. In the following text we will refer to spectra obtained from specific exposures, and we will use the shorthand notation ExpNN to refer to exposure number NN.

The data-set used here is a level-2 one downloaded from the Hinode Science Data Center Europe (\url{http://sdc.uio.no}) and was processed with version~1.83 of the IRIS level 1 to level~1.5 pipeline. Spectra derived from each exposure are averaged across a few pixels along the slit and version~4 of the IRIS radiometric calibration was applied. Gaussian profiles were fit to the emission lines in the spectra using the IDL routine \verb|spec_gauss_iris|. Wavelength calibration was performed by measuring the  \ion{O}{i} \lam1355.60 and \ion{S}{i} 1401.51 lines and assuming they are at rest in the spectra. The offsets were then applied to all other lines in the FUV1 and FUV2 channels. To derive Doppler shifts, the rest wavelengths  were taken from the line list available at \url{http://pyoung.org/iris/iris_line_list.pdf}.

Our first test of the QS-DEM method is to consider a compact  brightening seen in Exp38 (21:02:45 UT). From the SJI 1330~\AA\ images the brightening appears to belong to a loop, although this is spatially aligned almost exactly to the flare ribbon that appears about 10~minutes later. Thus the feature may represent early energy input to the flare ribbon. The brightening lasts for about 2~minutes and the peak intensity occurs in Exp36, but \lam1402.77 is saturated in this and the following exposure. For Exp38, pixels 225 to 227 along the slit were averaged, and Gaussian fit parameters for the emission lines are given in Table~\ref{tbl.ints}. The observed \ion{O}{iv} \lam1399.78/\lam1401.16 ratio of $0.394\pm 0.016$ is just below the high density limit and yields a density of $\log\,N_{\rm e}^{\rm Si\,IV}=11.96^{+0.35}_{-0.22}$ using data from Table~\ref{tbl.data}, while  \lam1402.77/\lam1401.16 is $32.4\pm 0.5$, indicating $\log\,N_{\rm e}^{\rm Si\,IV}=12.13\pm 0.01$. The results in  Table~\ref{tbl.ints} may also be used to predict that  (\ion{S}{iv} \lam1404.83 $+$ \ion{O}{iv} \lam1404.78)/\lam1401.16 should be 0.319 at $\log\,N_{\rm e}^{\rm Si\,IV}=12.0$, which is close to the observed ratio of $0.36\pm 0.01$. (Note that the uncertainties quoted here are the statistical uncertainties arising from the Gaussian fitting, and do not include atomic data, element abundance and DEM uncertainties which are likely to be at least 20\%.) We conclude that the QS-DEM method for computing the theoretical \lam1402.77/\lam1401.16 ratio thus provides good agreement with observed line intensities for this high density feature.

\begin{figure*}[t]
\epsscale{1.0}
\plotone{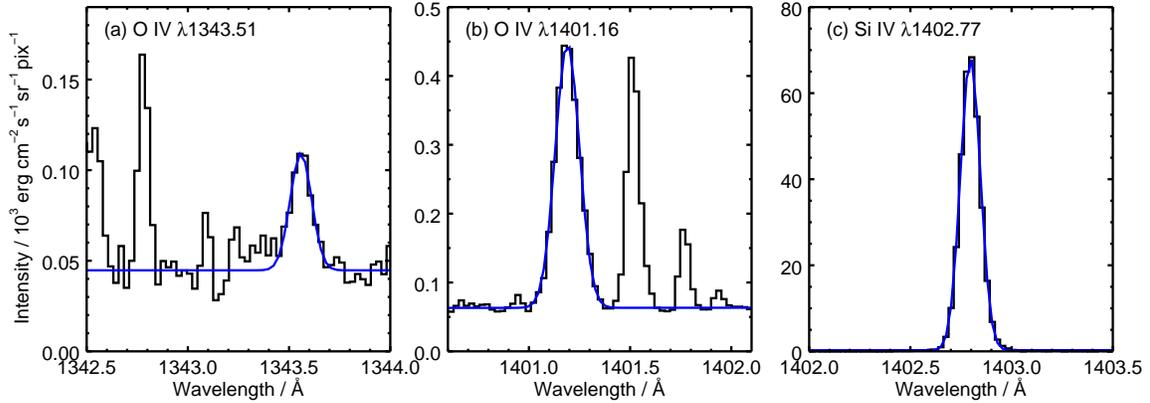}
\caption{Flare kernel spectra and Gaussian fits for Exp69 (panels a and b) and Exp70 (panel c).}
\label{fig.spec}
\end{figure*}

Now we consider flare kernel spectra from exposures 69 and 70, which mark the onset of exposure control. Exp69 (21:11:08~UT)  has a 15~second exposure time and clearly displays the \ion{O}{iv} lines, including \lam1343.51, but \ion{Si}{iv} \lam1402.77 is saturated, preventing an accurate intensity measurement (see Appendix~\ref{app.sat}). The exposure time drops to 0.44~seconds for Exp70 (21:11:18~UT) and  an unsaturated \ion{Si}{iv} line is detected together with very weak \ion{O}{iv} \lam1401.16, but \ion{O}{iv} \lam1343.51 can not be identified. Gaussian fits were performed to the \ion{O}{iv} lines from Exp69, plus \ion{O}{iv} \lam1401.16 and  \ion{Si}{iv} \lam1402.77 from Exp70 (Figure~\ref{fig.spec}), while the blend of \ion{O}{iv} \lam1404.78 and \ion{S}{iv} \lam1404.83 is also measured. For Exp69,   pixels 219 to 221 along the slit  were averaged, and 218 to 220 for Exp 70, the difference due to the rapid southward motion of the flare ribbon. 
Gaussians were fit to each line, and Table~\ref{tbl.ints} gives the 
centroids, line-of-sight velocities, full-widths at half-maximum (FWHMs) and integrated intensities. Due to uncertainties in establishing the spectrum background for \ion{O}{iv} \lam1343.51, the intensity error was increased by $\sqrt{2}$.

The weak \ion{O}{iv} \lam1401.16 line in the Exp70 spectrum was used to provide a correction factor for \ion{Si}{iv} \lam1402.77
as this is saturated in Exp69; the \ion{O}{iv} intensity was found to have changed by a factor $0.75 \pm 0.18$ between Exp 69 and Exp70. Hence we divide the Exp70 1402.77 intensity by this factor (see Table~\ref{tbl.ints}) to obtain a value that may be directly compared with the \ion{O}{iv} Exp69 measurements. However, we note that the error bars become rather large.  

The \ion{O}{iv} \lam1399.78/\lam1401.16 observed ratio is $0.47\pm 0.02$, which is greater than the high density limit (Table~\ref{tbl.data}). Thus we are in the high density regime where the diagnostic is no longer useful.
The \lam1402.77/\lam1401.16 ratio is $195\pm 62$, which yields $\log\,N_{\rm e}^{\rm Si\,IV}=12.96^{+0.09}_{-0.12}$  from Table~\ref{tbl.data}, while \lam1402.77/\lam1343.51  is $1327\pm 367$, significantly larger than the theoretical value from Table~\ref{tbl.data}. We also note that  (\ion{S}{iv} \lam1404.83 $+$ \ion{O}{iv} \lam1404.78)/(\ion{O}{iv} \lam1401.16) has a measured value of $0.78\pm 0.02$, yet the theoretical ratio is lower than this at all densities. For example, at $\log\,N_{\rm e}^{\rm Si\,IV}=13.0$ the ratio is 0.47. 
Since the intensity of \ion{O}{iv} \lam1404.78 is known to be consistent with the other \ion{O}{iv} lines \citep[e.g.,][]{keenan02}, the observed \ion{S}{iv} \lam1404.83 line must be about a factor of two stronger than the QS-DEM method predicts. 

The above results demonstrate some failings of the QS-DEM method for the high density regime. Hence the next step is to modify the DEM by making use of the extra information provided by \ion{O}{iv} \lam1343.51.

\begin{deluxetable*}{cccccccccccc}
\tablecaption{Emission line parameters. \label{tbl.ints}}
\tablehead{
  Ion & Line & Wavelength & Velocity & FWHM & Intensity & Exposure 
}
\startdata
\sidehead{Flare kernel}
\ion{Si}{iv}  & \lam1402.77 & $1402.795 \pm 0.009$ & $5.3\pm 2.0$ 
   & $24.6\pm 0.1$ & $328\,700 \pm 1500$ & 70 \\
 &  & \nodata & \nodata
   &  \nodata & $438\,800 \pm 104\,300$\tablenotemark{a} & 69 \\
\ion{O}{iv}  & \lam1343.51 & $1343.559\pm 0.010$ & $10.5\pm 2.2$ & $27.9\pm 2.8$ & $331\pm 47$ & 69 \\
 & \lam1399.78 & $1399.791\pm 0.010$ & $5.4\pm 2.0$ & $31.4\pm 1.3$ & $1051 \pm 44$ & 69 \\
 & \lam1401.16 & $1401.188\pm 0.009$ & $6.5\pm 2.0$ & $30.0\pm 0.3$ & $2251 \pm 26$ & 69 \\
 & & $1401.205\pm 0.017$ & $10.1\pm 3.6$ & $31.2\pm 7.4$ & $1686\pm 401$ & 70\\
 \ion{S}{iv} & \lam1404.81 & $1404.843\pm 0.009$ & $3.6\pm 2.0$ & $28.4\pm 0.4$ & $1754\pm 26$\tablenotemark{b} & 69 \\
 \sidehead{Bright point}
\ion{Si}{iv}  & \lam1402.77 & $1402.775 \pm 0.009$ & $1.1\pm 2.0$ & $17.3\pm 0.0$ 
   & $36\,069 \pm 84$ & 38 \\
\ion{O}{iv} 
& \lam1399.78 & $1399.784\pm 0.009$ & $3.9\pm 2.0$&$22.3\pm 0.8$ & $438\pm 17$ & 38 \\
& \lam1401.16 & $1401.170\pm 0.009$ & $2.5\pm 2.0$ & $22.4\pm 0.3$ & $1112 \pm 17$ & 38 \\
\ion{S}{iv} & \lam1404.83 & $1404.821\pm 0.009$ & $-1.0\pm 2.0$ & $23.7\pm 0.8$ & $401\pm 13$\tablenotemark{b} & 38\\
\enddata
\tablenotetext{a}{Intensity scaled from the Exp70 measurement---see main text.}
\tablenotetext{b}{Blended with \ion{O}{iv} \lam1404.78. Parameters derived for the blended feature.}
\end{deluxetable*}

\section{An updated diagnostic using O\,IV \lam1343}\label{sect.linlog}

Section~\ref{sect.qsdem} showed how a set of pressure--ratio curves could be derived for various combinations of \ion{O}{iv}, \ion{Si}{iv} and \ion{S}{iv} lines using only the assumptions of a fixed shape to the DEM and a particular set of element abundances. 
Comparing with line intensities from the 2014 October 24 flare kernel, we find two discrepancies: (1) \ion{O}{iv} \lam1343.51 is predicted to be too strong by about a factor two, and (2) \ion{S}{iv} \lam1404.81 is predicted to be too weak by a similar amount. Can we create a new model that fixes these discrepancies? 

Our procedure, which we refer to as the ``log--linear DEM method", is to define a new DEM by $\log\,\phi=a+b\log\,T$, and then use the measured intensities of  \lam1402.77, \lam1401.16 and \lam1343.51 to solve  Eq.~\ref{eq.int} for the three lines and yield  $P$, $a$ and $b$.
Using the flare kernel intensities from Table~\ref{tbl.data}, we derive $a=43.78$, $b=-3.90$ and $\log\,(P/{\rm K~cm}^{-3})=17.46$. The latter implies $\log\,N_{\rm e}^{\rm Si\,IV}=12.58$, which is 0.38~dex lower than that derived from the QS-DEM method.

If we now treat this DEM as a universal DEM then new pressure--ratio curves can be derived, and these are tabulated in Table~\ref{tbl.loglin}.
By definition, the new DEM fixes the problem with  \lam1343.51 noted for the QS-DEM method. We also find that the (\ion{S}{iv} \lam1404.83 $+$ \ion{O}{iv} \lam1404.78)/(\ion{O}{iv} \lam1401.16) ratio is predicted to be 0.69, significantly closer to the observed ratio of $0.78\pm 0.02$ than that from the QS-DEM method.

\begin{deluxetable*}{cccccccccccc}
\tablecaption{Theoretical ratios in energy units from the log--linear DEM method. \label{tbl.loglin}}
\tablehead{
   & \multicolumn{11}{c}{$\log\,(N_{\rm e}^{\rm Si\,IV}/{\rm
      cm}^{-3})$} \\
\cline{2-12}
  Ratio &
  9.0 & 9.5 & 10.0 & 10.5 & 11.0 & 11.5 &
  12.0 & 12.5 & 13.0 & 13.5 & 14.0 
}
\startdata
\sidehead{\ion{Si}{iv}/\ion{O}{iv} ratios}
\lam1393.76/\lam1401.16 &
   17.17&   17.20&   17.40&   19.28&   26.47&   48.60&  116.2&  325.7&  974.1& 2965.6& 8985.6\\
\lam1402.77/\lam1401.16 &
    8.61&    8.63&    8.73&    9.67&   13.28&   24.38&   58.31&  163.4&  488.7& 1487.8& 4506.6\\
\lam1402.77/\lam1343.51 &
  1478.4&  1530.2&  1563.2&  1557.0&  1497.6&  1413.8&  1354.6&  1326.1&  1308.4&  1282.7&  1224.4\\
\sidehead{\ion{O}{iv}/\ion{O}{iv} ratios}
\lam1401.16/\lam1343.51 &
 171.7& 177.4& 179.1& 161.0& 112.8&  57.98&  23.23&   8.115&   2.677&   0.862&   0.272\\
\lam1399.78/\lam1401.16 &
   0.172&   0.179&   0.197&   0.236&   0.299&   0.362&   0.401&   0.417&   0.423&   0.425&   0.425\\
\lam1404.78/\lam1401.16 &
   0.555&   0.494&   0.386&   0.274&   0.207&   0.180&   0.170&   0.167&   0.167&   0.166&   0.166\\
\sidehead{\ion{S}{iv}/\ion{O}{iv} ratios}
\lam1404.83/\lam1401.16 &
   0.051&   0.051&   0.052&   0.060&   0.087&   0.162&   0.315&   0.500&   0.623&   0.677&   0.699\\
\enddata
\end{deluxetable*}

\section{Application to flare kernels}\label{sect.kernels}

\begin{figure}[t]
\epsscale{1.0}
\plotone{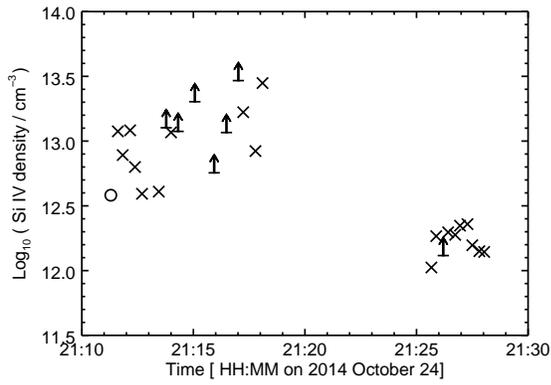}
\caption{Flare kernel densities for the 2014 October 24 data-set, derived with the log--linear DEM method. The circle corresponds to Exp70 (Table~\ref{tbl.ints}), and arrows indicate lower limits due to very weak \ion{O}{iv} emission.}
\label{fig.kernels}
\end{figure}

From the theoretical \lam1402.77/\lam1401.16 line ratio in Table~\ref{tbl.loglin} we can now derive densities for the two sequences of flare kernel measurements highlighted earlier (see Figure~\ref{fig.sji}). The first sequence occurred from 21:11 to 21:18~UT at approximately slit pixel 215, and the second from 21:25 to 21:28~UT at pixel 98. Kernels at the former position were around an order of magnitude more intense in the \ion{Si}{iv} line. For selected exposures, spectra were averaged over three to five pixels in the slit direction, and the two emission lines were fit with Gaussians. 
Only an upper limit to the \ion{O}{iv} intensity could be made for some exposures. 
Intensity ratios were then converted to $N_{\rm e}^{\rm Si\,IV}$ values using the data from Table~\ref{tbl.loglin}, and the results are plotted in Figure~\ref{fig.kernels}.

The kernels at the first position show a wide spread in density, which is mostly due to large variability in the \ion{Si}{iv} intensity, which has values ranging from $3.3 \times 10^5$~\ecss\ at Exp70 to $6.8\times 10^6$~\ecss\ at Exp75, with changes between exposures of up to a factor of of two 
(for example, between exposures 78 and 79). We note that during the exposure control period the FUV channel exposure times were $< 0.5$~seconds, while the average cadence was 16.2~seconds. Thus the IRIS intensities represent brief snapshots in the wider evolution of the flare ribbon. The large intensity variation suggests that we are observing multiple, short-lived energy input events at the spatial location rather than the evolution of single event.  We note that recent modeling efforts for flare kernels have suggested they consist of many structures below the resolution limit of IRIS \citep{2016ApJ...827..145R}.

The \ion{Si}{iv} intensity at the second location is  more stable and the corresponding density variation is smaller. This location is at the end of a hairpin shape in the flare ribbon, which can be seen in the bottom right corners of the middle and right panels of Figure~\ref{fig.sji}. The more uniform intensity suggests that the brightening may belong to a single energy input event, unlike the earlier observation.

\section{Summary and recommendation}\label{sect.summary}

The aim of the present work has been to create a model for the variations of ratios of \ion{Si}{iv} to \ion{O}{iv} emission lines in the IRIS spectra that will allow users to derive densities in the high density regime of $N_{\rm e} \ge 10^{12}$~cm$^{-3}$. Two approaches were adopted that differed in how the temperature structure in the atmosphere was modeled. The first was to assume that a quiet Sun DEM can be applied to all solar features, and checks against a spectrum for which the density is close to the high density limit of the \ion{O}{iv} \lam1399.78/\lam1401.16 ratio showed good agreement. However, the method failed to reproduce the strength of the \ion{O}{iv} \lam1343.51 line and the \ion{O}{iv} \lam1404.78 $+$ \ion{S}{iv} \lam1404.81 blend in a higher density flare kernel spectrum.

A second method, referred to as the log--linear DEM method, made use of \ion{O}{iv} \lam1343.51 in combination with \ion{Si}{iv} \lam1402.77 and \ion{O}{iv} \lam1401.16 to define a DEM that is linear in $\log\,T$--$\log\,\phi$ space over the region of formation of the ions. We took advantage of a special flare kernel data-set for which the exposure time dropped by a factor of 34 between two exposures, allowing good measurements of all three emission lines. This method yields lower densities than the QS-DEM one by about 0.4~dex, and is also better able to reproduce the intensity of the \ion{O}{iv} \lam1404.78 $+$ \ion{S}{iv} \lam1404.81 blend. Our recommendation is that the ratio curves from the log--linear DEM method (tabulated in Table~\ref{tbl.loglin}) be treated as universal curves to be applied for any case where the \ion{O}{iv} density diagnostic has reached its high density limit. For densities below this, the QS-DEM method seems to be more appropriate and has the advantage that it tends to the correct quiet Sun density value.

The log--linear method was then applied to two sequences of flare ribbon observations from the 2014 October 24 X-flare, and densities were derived. At one location of intense, variable activity the densities were found to lie between $10^{12.5}$ and $10^{13.5}$~cm$^{-3}$. A second location with weaker, more steady activity showed densities of around $10^{12.3}$~cm$^{-3}$. We note that these densities were derived assuming a constant pressure atmosphere and apply at the temperature of $\log\,T=4.88$. 

It is important to caution that the two diagnostic methods are best used for flare kernels and UV bursts. We know of at least one example of a structure for which the two methods will fail, and this is the sunspot plume. These structures have been studied by \citet{straus15} and \citet{2016A&A...587A..20C} and they can show \ion{O}{iv} \lam1401.16 intensities that are comparable to \ion{Si}{iv} \lam1402.77 and are thus not compatible with the ratio curves from Tables~\ref{tbl.data} and \ref{tbl.loglin}. This is likely due to a quite different temperature structure, and possibly also abundance anomalies.

We also caution that the methods are not intended to  be used for high precision density measurements. A simple translation of measurement errors to density uncertainties will generally yield very small errors on the density, but there are very significant uncertainties associated with the assumptions of the method: atomic data, the shape of the DEM and element abundances. A more realistic uncertainty is probably a factor of two. However, the ratio curves do provide a baseline for comparing different types of feature seen in IRIS data.

\acknowledgments

P.R.~Young acknowledges funding from NASA grant NNX15AF48G, and F.P.~Keenan is grateful
to the Science and Technology Facilities Council (STFC) of the UK for
financial support. R.O.~Milligan is grateful for financial support from NASA LWS/SDO Data Analysis grant NNX14AE07G, and to the STFC for the award of an Ernest Rutherford Fellowship (ST/N004981/1). P.R.~Young and H.~Peter thank ISSI Bern for support for the team ``Solar UV
bursts -- a new insight to magnetic reconnection''. IRIS is a NASA
small explorer mission developed and operated by LMSAL with mission
operations executed at NASA Ames Research center and major
contributions to downlink communications funded by the Norwegian Space
Center (NSC, Norway) through an ESA PRODEX contract.  CHIANTI is a collaborative project involving George Mason University, the University of Michigan (USA) and the University of Cambridge (UK). 

\facilities{IRIS} 
\software{Solarsoft, CHIANTI}

\bibliographystyle{aasjournal}
\bibliography{ms}

\appendix

\section{Saturated pixels for the FUV spectral channel}\label{app.sat}

As noted in the main text, the \ion{Si}{iv} lines often become saturated during observations of flare kernels and UV bursts. For density diagnostics we simply need the integrated intensity of an emission line. Can an accurate intensity be recovered from the saturated line profiles?

Saturation occurs when the signal from the CCD, in terms of numbers of electrons, is too high for the analog-to-digital convertor (ADC). For the IRIS cameras the saturation limit is 16\,283~DN. The gain for the FUV camera is set to 6~electrons~DN$^{-1}$, and thus a signal of 98\,000~electrons from a single pixel will reach the ADC saturation threshold.

The IRIS CCDs have a full well of 150\,000~electrons \citep{iris}, and therefore a photon flux for a single pixel that results in an electron count between 98\,000 and 150\,000~electrons cannot be accurately counted by the ADC. Above 150\,000~electrons, blooming will occur, whereby charge spills into neighboring pixels. If the neighboring pixels remain below the 98\,000~electron threshold, then the overspill charge can be accurately measured.

Note that if pixel binning is performed by the camera, then the situation becomes worse. For example, for $2\times 2$ binning (which was used for the 2014 October 24 flare event studied here) the ADC threshold for the binned pixel remains at 98\,000~electrons, whereas the effective full well capacity becomes 600\,000~electrons. Thus any signal between these two numbers is effectively lost.

To summarize, summing the intensity across the profile of a saturated emission line will likely significantly underestimate the actual intensity of the line. 

\end{document}